\documentclass[conference,a4paper]{IEEEtran}

\usepackage{amsmath}
\usepackage{amssymb}
\usepackage{xcolor}
\usepackage{amsthm}
\usepackage{mathtools}
\usepackage{siunitx}
\usepackage{cuted}
\usepackage{graphicx}
\usepackage{subcaption}
\usepackage{epstopdf}
\usepackage{tabularx}
\usepackage{multirow}
\usepackage{booktabs}
\usepackage{diagbox}
\usepackage{cuted}
\usepackage{lipsum}
\usepackage[noadjust]{cite}

\usepackage[english]{babel}
\usepackage[utf8]{inputenc}
\usepackage{algorithm}
\usepackage{algpseudocode}

\makeatletter
\newcommand{\removelatexerror} {\let\@latex@error\@gobble}
\makeatother

\newif\iftrackrivision

\iftrackrivision
	
\else
	
\fi

\IEEEoverridecommandlockouts

\begin{document}

\title{Multi-Sensory HMI for Human-Centric Industrial Digital Twins: A 6G Vision of Future Industry}

\author{
	\IEEEauthorblockN{Bin~Han~and~Hans~D.~Schotten}
	\IEEEauthorblockA{
	Technische Universit\"at Kaiserslautern\\
	\{bin.han~$\vert$~schotten\}@eit.uni-kl.de
	}%
}


\maketitle

\begin{abstract}
The next revolution of industry will turn the industries as well as the entire society into a human-centric shape. The human presence in industrial environment and the human participation in industrial processes will be magnified more than ever before. To cope with the emerging challenges raised by this revolution, 6G ambitions to bridge the three domains of digital information, physical assets and humans into one merged cyber-physical-human world. This proposes not only an unprecedented demand for digital twin solutions, but also new technical requirements. Especially, aiming at a human-centric industrial DT system, novel multi-sensory human-machine interfaces will play a key role in this paradigm shift.
\end{abstract}


\IEEEpeerreviewmaketitle

\section{Introduction}\label{sec:intro}
The concept of digital twin (DT) has been attracting profound attentions recently. It was first coined in 2014 in the context of manufactory to denote a virtual equivalent that models a physical product~\cite{grieves2014digital}. Triggered by the vigorous development of Internet-of-Things (IoT) and cloud computing technologies, the concept has been widely evolved and generalized over the past few years. Now it is commonly used to advocate a generic creation and maintenance of a virtual counterpart (the \emph{digital twin}) for arbitrary physical object or process (its corresponding \emph{physical twin}, or PT). Such a DT shall be capable to represent the status, the features, and the behavior of its PT in real time with the possibly finest accuracy. 

Such an image inevitably recalls the concept of cyber-physical systems (CPS), which refers to the systems that integrate physical units and processes with software and communication to provide abstractions and techniques for the integrated whole~\cite{shi2011survey}. Actually, any DT system is essentially an indivisible part of a CPS, which enhances the cyber-layer performance with advanced features such as real-time synchronization and high-level information construct~\cite{josifovska2019refernce}. Sometimes, the concept of DT system is also used in a wider sense to denote a subset of CPS, which have DTs implemented on its cyber layer \cite{kan2019digital}. Regardless the accurate definition, DT is generally believed to revolutionize CPS into its next phase, flourishing and emerging into the latest Fifth Generation (5G) mobile communication systems, not only with an enhancement to all 5G use cases, but also for the conveniences it brings to the test and validation of the 5G network itself~\cite{nguyen2021digital}. DT is also widely recognized to have a key role to play in the future Sixth Generation (6G) networks, especially in the industrial applications~\cite{wu2021digital,viswannathan2020communications,uusitalo2021hexa,uusitalo20216g,aiatedge2021d21}.

Despite the immature conceptualization of 6G, and the debate among different technical proposals, there is one feature of 6G that we know for sure to distinguish it from 5G: the human-centric orientation~\cite{uusitalo2021hexa,dang2020what,saad2020vision,zhang20196g,khan20206g}.

So far till now, the prosperous efforts in DT technologies have been focusing on - as its current definition suggests - bridging the gap between the physical world and the digital world, and have made a great achievement. To pave the way towards a human-centric 6G system, however, we are now challenged by the barriers between the cyber-physical world and the biological world. Reliable and affordable solutions are called for, to enable not only an immersive multi-sense human perception of the virtual world, but also an inclusion of human beings into the objective field of digital twinning.

In this survey, we outline the role of DT in future human-centric industrial 6G networking scenarios. As our main contribution that distinguishes this work from existing surveys on DT such as \cite{barricelli2019survey} and \cite{lim2020state,minerva2020digital,wu2021digital,wilhelm2020improving}, we focus on the identification of new challenges induced by the presence and participation of human in industry, the technical requirements and potential solutions to them. We also review the enabling human-machine interface (HMI) technologies, trying to shed the light on possible technical routes to break the human-CPS barriers.

The remainder of this paper is organized as follows: In Sec.~\ref{sec:dt_for_6g_i4p0} we throw a quick glance at the research trend, the application fields, and the use cases of DT. First in a generic view, and then in the context of future industry, identifying the gaps that cannot be fully closed by conventional 5G solutions but have to rely on 6G. Realizing the essential presence of human in the Industry 4.0 (I4.0) scenario of 6G era, in Sec.~\ref{sec:challenges_and_req} we try to identify the challenges of human-centric DT services, and the technical requirements that must be fulfilled. To overcome these challenges and seamlessly integrate the human participators into the DT-driven industrial practices, in Sec.~\ref{sec:hmi_survey} we review the cutting-edge HMI technologies, including human mental status sensing and multi-sensory feedback solutions, attempting to reveal some potential directions for future DT research. To the end, we close the paper with our conclusions in Sec.~\ref{sec:conclusion}.

\section{Digital Twin for 6G-Empowered I4.0}\label{sec:dt_for_6g_i4p0}

\subsection{Digital Twin: Concepts and Applications}
Research interest on the topic of DT has been explosively increasing over the past decade, spreading into various industrial branches, mainly including aviation, healthcare, and manufacturing. As a comprehensive survey~\cite{barricelli2019survey} shows, the early works were dominantly concentrating on aviation, where the expense of field experiments usually goes beyond affordability, which has encouraged the first burst of research efforts on DT. Since 2016, however, the focus has overwhelming shifted to the applications of manufacturing and precision medicine.

Despite the common vague opinion that each DT represents its corresponding PT, a universal agreement on the definition of DT is still absent. As pointed out by \cite{barricelli2019survey}: out of the 75 papers it reviewed, 44 did not give any explicit definition for the concept, with the rest 31 providing 29 different versions. Taking one more step beyond this analysis, we can further observe that among the 31 papers that explicitly define DT, 23 are focusing on the field of smart manufacturing, but highlighting various key points including integrated system, cyber-cloning, information connections, comprehensive digital construct, simulation test, and virtual replica. Through this superficial mist of a divergence (or even chaos) in opinions about the very same term, we see in DT a wild force driving the modern industry, especially its manufacturing sector, into the new era of future industry. In this paper, we generally refer as DT \emph{a digital replica of a living or non-living entity, whose virtual representation reflects all the relevant dynamics, characteristics, critical components and important properties of the original entity throughout its life cycle}~\cite{daria2021expanded}.

\subsection{The 6G Ambition of DT-Driven Industry}
I4.0, conceptualized in 2011~\cite{kagermann2011industrie}, seeks to digitize and automate the traditional manufacturing and industrial practices with mordern smart technologies. Identifying I4.0 as one of its key deployment scenarios, 5G has introduced a gallery of new use cases and technologies for support: the massive machine-type communications (mMTC) to enable dense IoT connections, the ultra reliable low-latency communications (URLLC) to guarantee timely and reliable data synchronization, the multi-access edge computing (MEC) to improve accessibility to artificial intelligence, the non-public network architecture to strengthen the security, and the vertical model to deliver heterogeneous services that are tailored to specific industry.

Now, while the deployment of 5G is rapidly spreading over the world, new ambition towards the next generation mobile networks for 2030s have already motivated up-springing research efforts, which aim at pushing everything beyond what 5G promises - and I4.0 cannot be missed. A new concept called Society 5.0 was proposed by the government of Japan, which extends the boarder of Industry 4.0 by putting human into the loop of CPS~\cite{fukuyama2018society}. Coinciding with each other on the human-centric view, Society 5.0 and 6G turn out to be a match made in heaven \cite{maier20216g}, leading to a pervasive CPS environment with ubiquitous connectivity - not only between everybody, between everything, but also between people and things. This implies to introduce a plethora of emerging human-centric industrial use cases to further enhance I4.0. According to the on-going European 6G flagship project \emph{Hexa-X}~\cite{uusitalo2021hexa,daria2021expanded}, 6G use cases can be clustered into several use case families:

\subsubsection{Massive twinning}
where the concept of DT is more fundamentally applied towards a full digital representation, extending the objective of DT to the entire environment, including both the physical and human worlds. This use case family contains not only new DT use cases such as immersive smart city and sustainable food production, but also a deeper integration of DT to the manufacturing environment. The mMTC scenario defined by 5G has established a solid basis of dense connectivity for this ambition. However, to practically realize the paradigm of massive twinning, it still lacks not only a richness of computing power at the network edge, but also reliable solutions to provision huge amount of data that drives the modeling, and global coverage to guarantee the continuous synchronization between DTs and their physical twins. These challenges cannot be fully addressed by the present 5G systems, but looking forwarding to 6G enablers.

\subsubsection{Immersive telepresence for enhanced interactions}
which enables humans to interact with all senses anytime anywhere, either with each other, with physical things, or with digital objects. This use case family opens a door to the future life, where the barrier of physical distance is completely removed for interactions in all fields of human society. The cyber-physical-human worlds will be fully merged by means of mixed reality (MR) and holographic telepresence technologies. In the industrial scenario, multi-sensory tele-interaction and telecollaboration will reshape the current I4.0 vision, which is based on 5G that cannot well support industrial MR application due to its limited throughput in URLLC and the lack of multi-sensory user interface.

\subsubsection{From robots to cobots}
differing from traditional robotic systems that are based on the command-and-control logos, future collaborative robots, a.k.a. ``cobots'', shall have relations not only with each other but also with humans. This new symbiotic relation will allow cobot systems to perform complex tasks in a collaborative and cognitive fashion. In the consumer electronics market, companion cobots are expected to better understand and serve the human needs than the off-the-shelf robots today; combined with the emerging AI-as-a-Service (AIaaS)~\cite{lins2021artificial}, personal AI agents that accompany with humans as autonomous working assistants or even equal partners are appearing no more out of reach. In the industrial environment, the flexible and intelligent collaboration among cobots and humans exhibits to us a bright prospect of flexible manufacturing. However, between this ambition and the state of the art, there is still a wide technical gap of unperceivable sensing and wireless AIaaS remaining open.

\subsubsection{Local trust zones for human and machine}
which aims at protecting privacy by means of dynamically updating the transparency and accessibility of sensitive information that are specific about individuals, machines, or independent networks, w.r.t. use requirements and privacy policies. It further evolves the concept of \emph{security trust zone (STZ)}, which was proposed in 5G regarding cellular topologies~\cite{han2017security,michalopoulos2018initial}, to make it apt to the beyond-cellular network topology that is not supported by 5G, e.g. the ``Network of Networks''. It grants trustworthiness to new use cases that are dense of sensitive information and rely on flexible network topology, such as precision healthcare, smart cities, public protection and disaster relief (PPDR), and automatic public security. In the industrial scenario, sensor infrastructure webs and low-power micro-networks will become the key use cases for production and manufacturing. 

\subsubsection{Sustainable development}
which attempts to address the social concerns about the sustainability of our environment and society, which has been globally growing over the past years, and especially intensified by the extreme weather events and disasters in 2021 (such as the European and Chinese floods, and the heats and wildfires in North America). Concerning more about the environmental and climate impact than 5G does, 6G takes sustainability as one of its key values. It is supposed to contribute to sustainability with new use cases that emphasize dematerialization, efficient resource usage, and energy-efficient optimizations, which all count in the industrial scenario. A deeper and meticulous understanding to the physical processes in industry is required hereby to improve the sustainability in a targeted manner.

In summary, we see in every 6G use case family a significant contribution to further evolve the I4.0. Each of them relies on an accurate, dynamic, real-time, and comprehensive modeling of all things, humans, and the environment - i.e. the upcoming 6G evolution to future industry is deeply rooted in the deployment of human-centric DT.

\section{DT with Human Presence in Future Industry:\\Challenges and Requirements}~\label{sec:challenges_and_req}
\subsection{Challenges by Human Existence in Industrial Environment}
\begin{figure}[!hbtp]
	\centering
	\includegraphics[width=\linewidth]{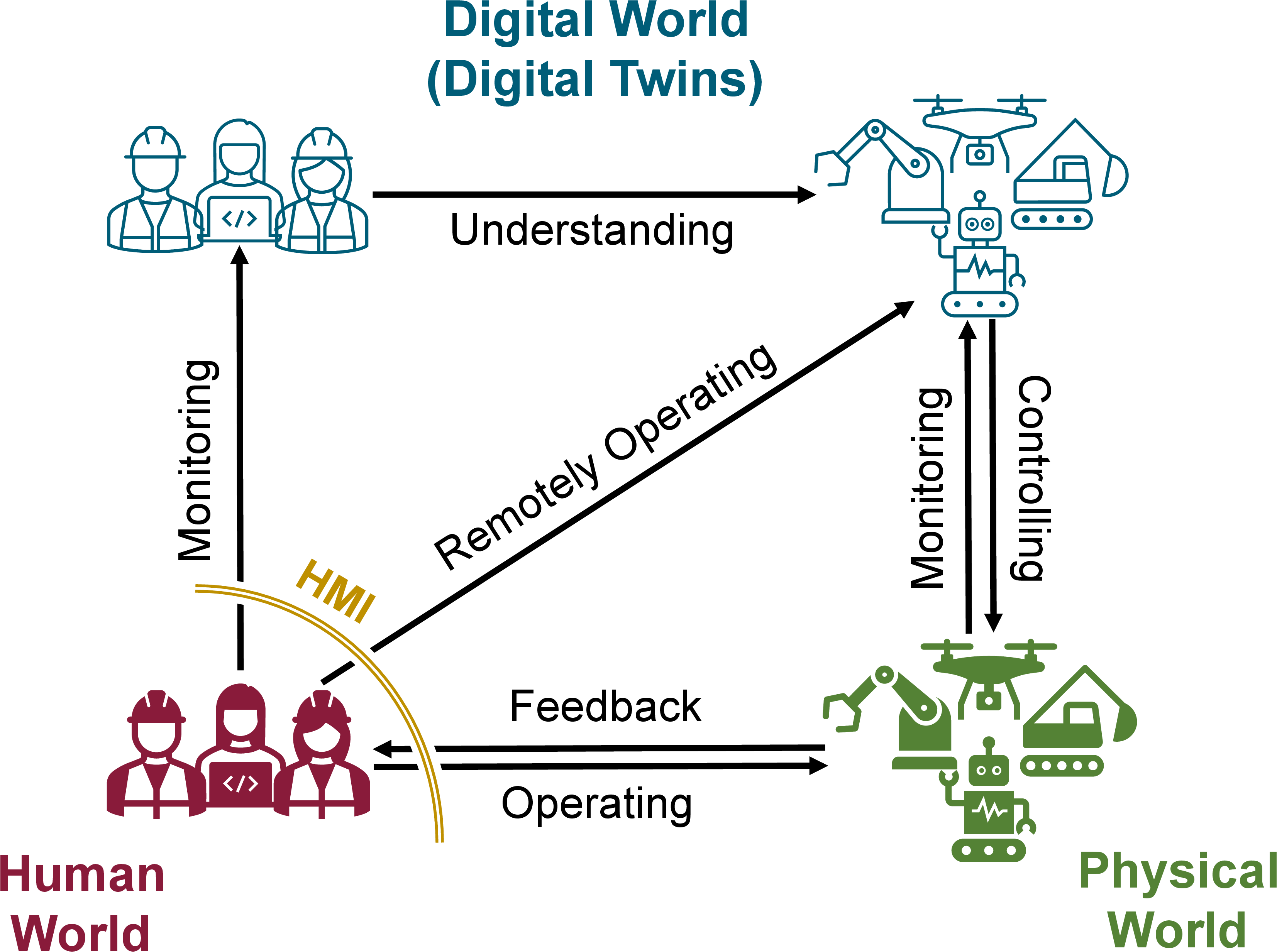}
	\caption{Information flow across the ``three worlds'' in future human-centric industry empowered by 6G DT. The arrows imply the directions of information flow.}
	\label{fig:dt_info_flow}
\end{figure}

These ambitious visions on future 6G-based industry are outlining an industrial DT ecosystem, where the human, equipment, as well as their DTs, converge and collaborate with each other, as illustrated in Fig.~\ref{fig:dt_info_flow}. Staying at the center of this ecosystem, humans rely on HMI technologies to interact with the other worlds.  So far, the direct interaction between human and physical equipment has been well supported by conventional technologies. However, upon the communication between humans and various DTs in the digital world, novel HMI solutions are required to efficiently address two classes of industrial challenges: the MR-collaboration and the human-disturbance protection.

On the one hand, the conscious participation of humans in industrial processes (such as controlling, manufacturing, and transporting) requires a dependable human-machine collaboration, both in onsite presence and over remote connection. This implies a huge demand in MR solutions based on multi-sensory interface. The traditional HMIs, which are dominantly based on mouse-keyboard-screen (MKS) and audio devices, shall be extended with holographic vision, haptic feedback, and unperceivable sensing, so as to allow human users to immersively interact with the cyber world, or interact with physical objects or other humans in telepresence over the DTs.

On the other hand, with their unpredictable actions, humans in industrial environment may usually, if not always, lead to disturbances or unexpected risky events. For example, pedestrians wandering in a factory can block the radio propagation in spectral bands of millimeter wave or higher frequencies, and therewith cause link failures of the wireless connected devices. For another instance, a human participator in industrial process may suffer from abnormal conditions such as sickness, fatigue, and strong emotions, which all reduce her level of concentration and comprehension, and thereby increase the risk of inappropriate operations that cause failures or dangers. To control and manage these risks raised by human existence, a timely monitoring of the human status - both physical and mental - and a thorough understanding of the complex human behavior, which are not sufficiently supported by conventional technologies, will be essential for the future industry.

\subsection{Technical Requirements for Novel HMI Solutions}
Indeed, the rapid development of HMI technologies have never been halted since the computer boom in the 1980's. A large variety of solutions have been invented, ranging from the gyroscope-accelerometer-based motion technologies that are already commonplace nowadays, to the brain computer interfacing (BCI) solutions that are well demonstrated~\cite{cannan2011human} but yet sounding science-fictional. However, not all of them are promising for the 6G-based future industry. More specifically, we identify the following technical requirements as essential for any 6G-HMI to drive the next industrial revolution.

\subsubsection{Safety and health-friendliness}
as a human-centric system, 6G takes the safety and health of humans as its core value and first priority. Axiomatically, any technology with potential risk of damaging the health of users by itself shall never be considered before being eventually verified as harmless. In addition to that, due to the particularity of industrial environments, every candidate HMI solution shall be evaluated in context of the specific use scenario, to make sure that it does not significantly distract or restrict the user from his main task. 

\subsubsection{Comfort and convenience}
beyond the basic requirements of health and safety, user's comfort and convenience shall be taken into account. This generally requires a compactness of the essential hardware and rejects most (if not all) invasive interfaces. For example, to recognize facial expressions and gaits, graphic solutions based on cameras \cite{kusakunniran2009automatic,cohen2003facial} are preferred in this sense than solutions based on electromyogram (EMG) signals that can only be measured with electrodes \cite{lai2009computational,gruebler2014design}; should the EMG signals be measured, surface electrodes are much preferred than needle electrodes.

\subsubsection{Dependability}
for applications intolerant to outages, 6G intends to deliver guarantees for multiple end-to-end (E2E) performances, such as achievable data rate, maximum E2E service latency, bounded jitter, and E2E packet reliability with robust mobility. This is known as the concept of \emph{dependability}~\cite{uusitalo2021hexa}, which is especially critical for use cases such as human-machine interaction and automation. As an essential stage of the E2E service chain in 6G-driven industrial DT, the HMI must cope with these requirements.

\subsubsection{Electromagnetic Compatibility}
massive twinning in industrial scenario inevitably leads to a complex electromagnetic environment, where strong interference and noise may probably occur, sometimes even randomly and without predictable pattern. The HMI solution therefore must be electromagnetically robust to ensure its reliability. On the other hand, it shall not generate strong electromagnetic leakages that interfere with the wireless channels or other devices.

\subsubsection{Sustainability}
6G takes sustainability also as one of its key values. This does not only mean that 6G shall foster improved sustainability in various societal domains, but also implies that 6G itself must be made sustainable. Unsustainable technologies, which depend on nonrenewable resources, use pollutive materials, or generate high CO\textsubscript{2} submission, shall not be adopted in 6G - with no exception for the HMI.

\subsubsection{Security and privacy}
With its ubiquitous coverage and massive twinning, 6G will connect everything and everybody with each other, carrying massive data to comprehensively describe them in more details than ever before. However, it also raises security and privacy concerns to an unprecedented level. On the one hand, enhanced security measures must be taken to guard the user data from unauthorized access and malicious operation by the non-trusted. On the other hand, the user data also must be protected from inappropriate exploitation by the trusted such like the industrial verticals. For instance, the General Data Protection Regulation (GDPR) of European Union prohibits a handful kinds of data processing that can leak the user identity or lead to discrimination,
with only a few exceptions under very strict rules. How to exploit human-specific data in DT systems (especially those with DTs of humans), while staying aligned with such regulations, must be taken into consideration when designing the HMI.

\section{HMIs for Human-Centric DT: An Overview}\label{sec:hmi_survey}
Like we have discussed above, technical challenges for future industrial DT systems with human presence are identified in both aspects of HMI: the sensorologic one of sensing and monitoring of human status from the machine side, as well as the perceptual one of generating feedback in MR environment to humans. On the sensorologic side, mature solutions to sense and monitor the mobility and physical status (e.g. body temperature, pulse rate, etc.) have been well developed and widely applied in commercial personal devices over the past decade, while the sensing of mental status is remaining as an open issue. On the perceptual side, conventional technologies are still generally relying on visual and auditory senses, leaving the other human senses (tactile, olfactory, and gustatory) barely exploited. To inspire innovations towards an immersive human-centric industrial DT, we give here an overview to the potential technical enablers of human mental status sensing and multi-sensory feedback.

\subsection{Sensing the Mental Status}\label{subsec:sensing_mental_status}
Industrial processes can be greatly impacted by the mental status of the human participators, not only regarding the quality of output/product, but more importantly, regarding the safety of system and humans themselves. The most critical mental signals to be measured include:

\subsubsection{Comprehension} in industrial scenarios, especially regarding human-machine collaboration, it is usually important for the intelligent systems to confirm that the human has perceived and understood the critical information it provides, such like a notification, an inquiry, an instruction, or a warning.

\subsubsection{Concentration} distraction from the task in processing in industrial environment can easily lead to operation failures, which may cause serious losses and dangers.

\subsubsection{Fatigue} fatigue generally reduces the level of concentration. Besides, it also weaken human's strength, agility, precision, and endurance in physical tasks; and degrades the human cognitive abilities in all aspects. In addition, continuous working under fatigue is likely to damage the health.

\subsubsection{Emotion} negative emotions such as depression, stress, anxiety and anger can regularly lead to distraction, and accelerates the increase of fatigue. With the significant impact on the human behavior pattern, extreme emotions may also cause failures of human-related prediction algorithms.

Since direct sensing of the mental status is impossible, people have developed a handful of indirect approaches to estimate it from measurable physical features. Such physical features that are typically taken as indicators include:

\setcounter{subsubsection}{0}

\subsubsection{Speech voice} emotions and fatigue can significantly change the tone, speed, volume and timbre of one's speech voice. Hence, they can be detected through vocal analysis. However, to continuously monitor the mental status of a person in this way, the person has to keep talking, which is not universally possible. Nevertheless, in some certain scenarios where the vocal signal is available, e.g. when vocal commands are used to control the intelligent system, voice-based mental status estimation can be well integrated.

\subsubsection{Facial expressions} rich information about one's mental status can be mined from her facial expressions and microexpressions, which has been exploited by human beings for thousands of centuries as most classical and reliable approach of emotion recognition. This approach requires images or videos that capture most facial areas - in sufficiently high resolution and ideally in the front view - of the person under analysis, which are easy to obtain in most industrial scenarios.

\subsubsection{Galvanic skin response signal} shifting in the intensity of emotional states can cause effects in human's eccrine sweat gland activity, which change the conductance of skin. Such effects are almost immediate to the emotional arousal, and easy to measure with wearable devices.

\subsubsection{Eye movements} eye movement have been used since long as a psychological signal for emotion recognition. It can be captured either from video by cameras and eye trackers, or from electrooculogram signals by specialized electrodes. Both can be easily integrated into wearable devices such like VR/AR headsets.

\subsubsection{Bioelectric signals} there are a variety of bioelectric signals widely used in psychological and clinical studies to identify human brain activities and mental status, common examples include electroencephalogram (EEG), electromyogram (EMG), electrooculogram (EOG), and electrocardiogram (ECG). They usually need to be sensored with specialized electrodes, which are not always convenient and comfort to carry while working in the industrial environment.

Some examples of mental status sensing are listed in Tab.~\ref{tab:sensing_mental_status}.

\begin{table}[!hbtp]
	\centering
	\caption{Examples of mental status estimation with novel HMIs}
	\label{tab:sensing_mental_status}
	\begin{tabular}{l|l|l}
		\toprule[2px]
		\textbf{Mental signal}			&\textbf{Physical feature}	&\textbf{Literature}\\
		\midrule[1.5px]
		\multirow{4}{*}{Comprehension}	&Facial expression			&\cite{turan2018facial}\\
										&Galvanic skin response		&\cite{bottos2020tracking}\\
										&Eye movements				&\cite{li2016your}\\
										&Bioelectric signals 		&\cite{rusak2016smart}\\
		\hline
		\multirow{2}{*}{Concentration}	&Facial expression			&\cite{sharma2019student,takahashi2005method}\\
										&Bioelectric signals 		&\cite{kaji2019ecg}
		\\\hline
		\multirow{5}{*}{Fatigue}		&Speech voice				&\cite{greeley2006detecting}\\
										&Facial expression			&\cite{khan2018effective}\\
										&Galvanic skin response		&\cite{yang2005driver}\\
										&Eye movements				&\cite{singh2010eye}\\
										&Bioelectric signals 		&\cite{kolodziej2020fatigue}\\
		\hline
		\multirow{5}{*}{Emotion}		&Speech voice				&\cite{huang2014speech}\\
										&Facial expression			&\cite{liu2017facial}\\
										&Galvanic skin response		&\cite{liu2016human,susanto2020emotion}\\
										&Eye movements				&\cite{soundariya2017eye}\\
										&Bioelectric signals 	 	&\cite{subramanian2018ascertain}\\
		\bottomrule[2px]							
	\end{tabular}
\end{table}

\subsection{Multi-Sensory Feedback}
Since vision and hearing are the most important sense of humans, visual and auditory user interfaces (UIs) will still indefinitely remain the dominating approach for machines sending information to humans. However, the traditional visual/auditory UIs based on text, two-dimensional graphics and audio cannot fulfil the requirements of future industrial DT applications like MR, immersive telepresence, and human-robot collaboration, and therefore must be extended. 

\subsubsection{Holographic vision}
As the successor of text, image, video and 3D video, holography will be the fifth generation of visual user interface technology. Allowing such a visualization of DT, that no clear boundary is to be sensed between the virtual objects and the real physical environment, holographic vision plays a key role in the future DT applications including MR, immersive telepresence, and telecollaboration.

There have been already commercial holographic vision products released, which have also been applied in research works, such as the Microsoft HoloLens 2~\cite{ungureanu2020hololens}. They have been proven sufficient as solutions to optically deliver MR. To achieve the target of teleinteraction and telecollaboration, however, there are still two main challenges to overcome. First, the closed-loop latency must be minimized, taking into account the computing delay to update the holographic model of DT regarding human actions (e.g. pressing a button on the holographic projection of a machine), so as to realize a smooth teleinteraction experience. Second, when multiple users are involved with, e.g. in telecollaboration, the holographic image/video must be highly synchronized for all users. Both challenges are related to the research areas of URLLC, information freshness, and time-sensitive networking (TSN).

\subsubsection{Tactile}
Alongside with vision, the mostly used human sense, there is the tactile that is used by us in every physical contact of ours with the environment or any object. Being partly processed by the spinal cord instead of the brain, tactile is also the most agile sense of humans, responding significantly faster than the visual and auditory senses~\cite{ng2012finger}. Since the concept of \emph{Tactile Internet} being proposed in 2014 \cite{fettweis2014tactile}, the telecommunication community has been struggling to achieve the target of $\leqslant10$\si{\milli\second} E2E latency, which is the limit of tactile cognition for humans. In the wireless field, the 5G URLLC use case was proposed to address this issue, breeding numerous research works over the past years. 

Despite the significant progress in evolving the network infrastructure to support tactile internet, haptic solutions to generate agile and accurate tactile feedback are still far behind the requirements of future industry. The haptic interface is a critical enabler for industrial DT applications, not only because it enriches the user experience of sensing virtual objects with tactile, but also as a necessary condition of reliable motion-capture-based remote operation. Only with an accurate and fast haptic stimulus that vividly simulates the weight, resistance and hardness of physical objects, it becomes possible for a human user to appropriately exert and accomplish high-precision tasks such like surgery over teleinteraction.

Traditional solutions to provide tactile feedback are mostly \emph{extrinsic}, i.e. they rely on instrumented environment with active devices, such like vibrators-embedded steering wheels for virtual automotive driving~\cite{diwischek2015tactile}, and the vibration-based tactile simulation of physical keyboard on ~\cite{brewster2007tactile}. Such solutions are generally limited to local area and specific use scenarios. To enable tactile feedback in immersive and ubiquitous teleinteraction with arbitrary object, the \emph{intrinsic} solutions try to augment the user instead of the environment, and alter the user's tactile perception. This can be typically realized by wearable haptic devices, or through direct stimulation to the user's neurosensory mechanism~\cite{bau2012revel}. Recently, there has been a new class of methods developed, which use drones to flexibly generate tactile feedbacks~\cite{abdullah2017haptic,knierim2017tactile}.

\subsubsection{Other senses}
The senses of smell and taste may also be useful in some industrial scenarios, but the physicalization of them remains a technical challenge~\cite{jansen2015opportunities}.

\subsubsection{Spinal cord and brain stimulation}
Besides stimulating the natural human perception system, it is also possible to generate artificial sensory feedback by directly stimulating the human neural system, i.e. spinal cord and brain. This topic has been studied since long in the field of neuroscience, the results have been so far mainly applied in medical treatments and devices, such as prosthetic implants that can simulate tactile sense for amputees~\cite{tan2015stability}. This type of approaches can be outstanding in the effectiveness of triggering sense in flexible environments, but may suffer from high implementation cost and raise more concerns in human safety. Recently, it has become an emerging topic of combining such technologies with the future 6G wireless solution towards a so-called \emph{brain-type communications} scenario~\cite{moioli2021neurosciences}.

\subsubsection{Synesthesia}
Additionally, in certain scenarios it may be useful to leverage the synesthetic phenomena, wherewith one sense can be triggered by another that is easier to physicalize. For example, it has been reported that by watching/listening to certain type of videos/audios, it triggers a special gentle-touch-alike tactile sense accompanied by a deep relaxation and pleasure, which is known as the autonomous sensory meridian response (ASMR)~\cite{peorio2018more}. Such phenomena are suggesting a new class of solutions to resolve/soften the negative mental status we discussed in Sec.~\ref{subsec:sensing_mental_status}. To achieve this, deep study is still required to setup a reliable model of the human sensory system, where DT can also play an important role.

\section{Conclusion}\label{sec:conclusion}
In this paper we have provided our vision on the role played by DT in future 6G-empowered industry. We have identified the gaps, technical requirements and challenges for novel HMI solutions in such industrial DT applications, taking a human-centric point of view. We have also surveyed the main potential technology routes and candidate solutions, hoping to inspire interested peers towards the next breakthrough in this field.

\section*{Acknowledgement}
This work has been partly funded by the European Commission through the H2020 project Hexa-X (GA no. 101015956).

\ifCLASSOPTIONcaptionsoff
  \newpage
\fi


\end{document}